\documentclass[aps, twocolumn]{revtex4-1}
\usepackage[utf8x]{inputenc}
\usepackage{placeins} 
\usepackage{epsfig}
\usepackage{amsmath,latexsym}
\usepackage{color}
\usepackage{xcolor, soul}
\usepackage[toc,page]{appendix}
\usepackage{graphicx}% Include figure files
\usepackage{dcolumn}% Align table columns on decimal point
\usepackage{bm}% bold math
\usepackage[normalem]{ulem} %Strike through
\usepackage{textgreek}
\usepackage{enumitem}
\usepackage{pifont}

\begin{document}

\title{Microfluidics control the ballistic energy of thermocavitation liquid jets for needle-free injections}
\author{Loreto Oyarte G\'alvez$^{1,2}$} 
\author{Arjan Fraters$^{3}$}
\author{Herman L. Offerhaus$^{4}$}
\author{Michel Versluis$^{3}$}
\author{Ian W. Hunter$^{5}$}
\author{David Fern\'andez Rivas$^{1,5}$} \email{d.fernandezrivas@utwente.nl}
\affiliation{$^{1}$Mesoscale Chemical Systems Group, MESA+ Institute, TechMed Centre, and Faculty of Science and Technology, University of Twente, P.O. Box 217, 7500AE Enschede, Netherlands}
\affiliation{$^{2}$Current address: Department of Ecological Science, Faculty of Earth and Life Sciences, Vrije Universiteit Amsterdam, Amsterdam, 1081HV, The Netherlands}
\affiliation{$^{3}$Physics of Fluids Group, MESA+ Institute, TechMed Centre, University of Twente, P.O. Box 217, 7500AE Enschede, Netherlands}
\affiliation{$^{4}$Optical Science Group, MESA+ Institute and Faculty of Science and Technology, University of Twente, P.O. Box 217, 7500AE Enschede, Netherlands}
\affiliation{$^{5}$BioInstrumentation Laboratory, Department of Mechanical Engineering, Massachusetts Institute of Technology, Cambridge, MA 02139, USA}

\date{\today}
 
\begin{abstract} 
Illuminating a water solution with a focused continuous wave laser produces a strong local heating of the liquid that leads to the nucleation of bubbles, also known as thermocavitation.
During the growth of the bubble, the surrounding liquid is expelled from the constraining microfluidic channel through a  nozzle, creating a jet. The characteristics of the resulting liquid jet was imaged using ultra-fast imaging techniques. Here, we provide a phenomenological description of the jet shapes and velocities, and compare them with a Boundary Integral numerical model.  We define the parameter regime, varying jet speed, taper geometry and liquid volume, for optimal printing, injection and spray applications.
These results are important for the design of energy-efficient needle-free jet injectors based on microfluidic thermocavitation.

\end{abstract}

\maketitle
%%%%%%%%%%%%%%%%%%%%%%%%%%%%%%%%%%%%%%%%%%%%%%%%%%%%%%%%%%%%%%%%%%%%
%%%%%%%%%%%%%%%%%%%%%%%%%%%%%%%%%%%%%%%%%%%%%%%%%%%%%%%%%%%%%%%%%%%%
\section{Introduction}\label{s:Intro}
When laser light is focused in a liquid with a sufficiently high absorption of the particular laser wavelength, a vapour bubble can be formed~\cite{ohl_acoustic_2016, xiong_droplet_2015, shen_dynamics_2014, zhu_pancake_2016,Brujan,Wang201805912}. Depending on the confinement conditions this bubble can grow and expel the surrounding liquid through a nozzle creating a liquid jet. This laser-induced cavitation has been proposed for innovative jet printing~\cite{Moser,Moser2}, and needle-free injection~\cite{gonzalezavila2015,Tagawa2,Berrospe}. While the high energy pulsed-laser based systems produce fast jets, even at supersonic velocities, they are expensive, bulky and non-portable, among other drawbacks~\cite{Berrospe, padilla-martinez_optic_2014, Peters}. 

Bubbles made with continuous wave (CW) lasers, known as thermocavitation~\cite{rastopov1991sound}, may offer several advantages over systems using high energy pulsed-lasers~\cite{padilla-martinez_optic_2014}. CW diode lasers are becoming increasingly cost-effective, both high and medium power units (tens of Watts and less, respectively), and are used in single mode and multimode. The most abundant systems, for telecom and remote sensing applications, have near infrared (NIR) wavelengths. Visible diode lasers are also available, and these are considered safer for eye and skin, and can thus easily be incorporated into consumer products. The electrical power to light energy conversion can be as high as 20\% making them an efficient means to deliver energy to the fluid~\cite{coldren2012diode,pietrzak2013progress}. In addition, the difference in energy values used as compared to high energy pulsed lasers, the vapour bubble dynamics, as well as the jet velocities reached, show distinct geometrical features that we are just beginning to understand~\cite{sun_can_dijkink_lohse_prosperetti_2009, zeng_2018, brujan_2018}. 

Controlling the velocity, diameter, and shape of liquid jets, is crucial to produce fast-travelling liquid micro-droplets for industrial and biomedical applications, including ink-jet printing, cleaning, and jet injectors for drug delivery through the skin~\cite{hoath2016fundamentals,TIRELLA201179,hu2018functional,prausnitz2004current,verhaagen2016measuring,mitragotri2014overcoming,oyarte2019high,Cu}. The shape of the jet is of particular importance for inkjet printing, where a spherical jet tip and stable and reproducible drop formation are required~\cite{Hue,Park2007,Hongming}. Moreover, the jet speed together with the liquid viscosity limit the range of a printable fluid to an Ohnesorge number, {\it i.e.\ } the ratio of viscous dissipation to the surface tension energy, between Oh=0.1 and Oh=1~\cite{McKinley,MCILROY201317,Derby}.

CW lasers have only recently been employed for the generation of jets with cavitation inside microfluidic chips~\cite{Berrospe}. In a follow-up study, devices with different nozzle diameters in the range of 100~$\mu$m achieved maximum jet velocities of up to $\sim$~95~m/s, and the injection into soft hydrogels was recorded~\cite{Berrospe2}. Unlike jetting in cylindrical geometries, this device had a rectangular cross-section channel with high aspect ratio known to produce disk-like cavitation bubbles\cite{zhu_pancake_2016,Zwaan}, with different self-focusing effects as those observed in capillaries. The tapering angle of the nozzle was varied in a range of $15^\circ$ to $40^\circ$, and provided a speed increase of up to $65~\%$~\cite{Berrospe2}.

Jetting phenomena can be influenced by additives, e.g.\ polymers, that change liquid properties such as viscosity and elasticity~\cite{mcilroy2013modelling}. However, the control of these jets is difficult in practice because such additives change the better understood Newtonian liquids jetting conditions. The polymer type, concentration, liquid temperature and various non-linear phenomena, such as asymmetric jet formation and pinch-off, can in turn reduce the kinetic or ballistic energy of the resulting jets~\cite{ardekani_sharma_mckinley_2010, Onuki, yuen_1968, Nayfeh, rohilla2019characterization}. Besides manipulating the liquid properties, an elegant approach can be found in controlling the geometry of the microfluidic channel in which the vaporisation and jetting phenomena occur. The meniscus formed in such confined conditions can provide self-focusing of the energy during jet formation, and the jet diameter is less influenced by the nozzle geometry or capillary diameter~\cite{Tagawa}.

Small orifices used to deliver small jets have a higher probability of clogging, and typically require expensive fabrication techniques~\cite{grande2016direct,Shang:2017jy,dressaire2017clogging}. Additionally, the energy losses caused by flow through such small orifices lowers the kinetic energy of the jet to a point that either splashing or bouncing can occur~\cite{Gielen2018}. Interestingly, an increase in penetration depth takes place with increasing nozzle diameter at a constant exit velocity, and {\it vice versa} with increasing jetting velocity at constant diameter~\cite{BaxterMitragotri2004}. However, larger nozzle diameters mean lower jet pressures and therefore the delivered percent volume decreases leading to a reduction in the overall performance~\cite{schramm2002transdermal}. Two criteria can be considered for skin failure: 1) comparing the local normal stress induced by the jet impact with a critical local stress, and 2) comparing the energy density input to the skin with a critical energy density~\cite{schramm2002transdermal}. Independent from the exact values, both local stress and energy of the jet/skin system depend on the jet speed $U_\text{jet}$ and radius $r$. The local static stress of the jet $p_{jet}=1/2\rho v_\text{jet}^{2}$ acting on an area defined by the jet radius, will deliver its kinetic energy during the working time of the jet: $E_{k}= \pi r^{2} p_{jet}t$~\cite{schramm2002transdermal}. However, two phenomena strongly affect the injection process. One is that jets can break-up before impact and reduce the total volume injected before the hole or pore pierced in the substrate closes. The second is that upon impact at the substrate or skin, there is a splash-back that influences the efficacy of the payload delivery, and that increases contamination risks between subsequent injection events~\cite{arora2008micro}. In practice, the jet speed and the jet-tip shape are crucial parameters for the design of an efficient jet injector. It is accepted that jets of $\sim$20$~\mu$m in diameter with a jet speed of $\sim$15~m/s can puncture the skin~\cite{Berrospe2}.  Until now, jets made with CW lasers have attained modest velocity values compared to jets achieved with injector concepts such as pulsed lasers (850 m/s)~\cite{Tagawa} and voice coils (100 m/s)~\cite{mckeage2018effect}. However, as evidenced in a recent study, thermocavitation jets seem to perfuse ex-vivo porcine skin with jet velocities as low as 15 m/s, reaching depths in the range of 0.05-0.22 mm (with single and up to six injections in one spot~\cite{Cu}.

\begin{figure}[h!]
  \begin{center}
   \includegraphics[scale=1]{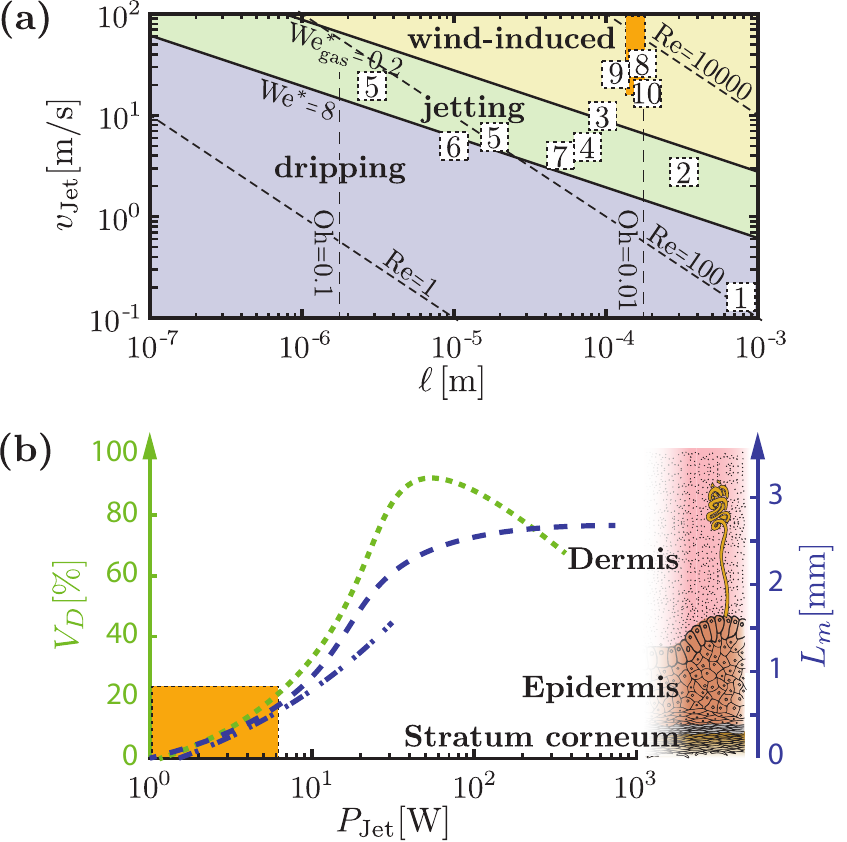}
  \end{center}
  \vspace{-0.3cm}
  \caption{(a) Liquid jet breakup regimes dependence on the characteristic length $\ell$ and jet speed $v_\text{Jet}$. The dimensionless parameters Re=$\frac{\rho\ell v_\text{Jet}}{\mu}$, We$^*=\frac{\rho\ell v_\text{Jet}^2}{\sigma}$, We$^*_\text{gas}=\frac{\rho_\text{air}}{\rho}\text{We}$ and Oh=$\frac{\mu}{\sqrt{\rho\sigma\ell}}$ are calculated for the aqueous solution used in this work. The boxed-numbers refer to the studies in jet formation performed by: (1) Ambravaneswaran \textit{et al.}~\cite{Ambravaneswaran}, (2) Kalaaji \textit{et al.}~\cite{Kalaaji}, (3) Gonz\'alez and Garc\'ia~\cite{gonzalez}, (4) Fainerman \textit{et al.}~\cite{FAINERMAN}, (5) Hoeve \textit{et al.}~\cite{Hoeve}, (6) Suk Oh \textit{et al.}~\cite{OH201427}, (7) Etzold \textit{et al.}~\cite{ETZOLD}, (8) Berrospe \textit{et al.}~\cite{Berrospe2}, (9) Cu \textit{et al.}~\cite{Cu} and (10) Oyarte Galvez \textit{et al.}~\cite{Oyarte}. (b) Delivered volume percentage $V_D$ (green) and penetration depth $L_m$ (blue) with respect to the jet power $P_\text{Jet}=\frac{1}{2}\rho Av^3_\text{Jet}$ in the case of needle-free injection in skin, the curves refer the work of: Schramm-Baxter \textit{et al.}~\cite{BaxterMitragotri2004} (dashed line) and Tagawa \textit{et al.}~\cite{Tagawa2} (dotted-dashed line). The orange rectangles, in (a) and (b), represent the regimes where the jets in this work are located. In the case of (b), we calculate the delivered volume using the experimental power values of this work and the theoretical approximation from Schramm-Baxter \textit{et al.}~\cite{BaxterMitragotri2004} for injection in real skin.}
  \label{fig:OurWork}
\end{figure}

The power of the jet is calculated as: 
\begin{equation}
P_\text{Jet}=\frac{1}{2}\dot{m}v_\text{Jet}^2=\frac{1}{2}\rho A v_\text{Jet}^3
\end{equation}
\noindent where $\dot{m}=\rho A v_\text{Jet}$ is the jet mass flux in time and $A$ is the nozzle cross-sectional area. The jet speeds achieved in our experiments were between $\sim$20~m/s and $\sim$100~m/s, $A=100\times120~\mu$m$^2$, and therefore the jet power is in the range [50~mW, 6~W]. The penetration depth $L_m$ and the delivered volume percentage $V_D$ of injection in the skin depend on the jet power, as we present in an adapted plot in Fig.~\ref{fig:OurWork}~(b), based on previous work~\cite{BaxterMitragotri2004,Tagawa2}. In the first case, injection experiments were performed in human skin using a commercial spring-driven jet injector (Vitajet 3, Bioject, Portland, OR). In the second case, highly-focused high speed microjets were injected into skin and soft matter. The orange rectangles cover our study, where the jets  deliver a volume $V_D\simeq25\%$ of the total ejected volume, and may reach an equivalent depth of up to $L_m\simeq1$~mm in human skin. The same jet injector reported in this study as shown that the injected volume efficiency in {\it ex vivo} porcine skin can be as high as 75--90\%~\cite{Cu}.

Here, by introducing CV microfluidic thermocavitation, we aim at increasing the volume delivered with the same or less power, and at the same time, establish a window of opportunities in the parameter space in which jets are created with an optimal geometry avoiding break up before reaching its target site.
%%The purpose of this paper is to improve the current understanding of the jet formation with thermocavitation and to determine how to maximise the energy transfer from the laser into kinetic energy of the jet. We achieve this through numerical modelling and fast-imaging experimental observations of the effect of taper angle variations on the characteristics of liquid jets at constant laser power. 

%%%%%%%%%%%%%%%%%%%%%%%%%%%%%%%%%%%%%%%%%%%%%%%%%%%%%%%%%%%%%%%%%%%%
%%%%%%%%%%%%%%%%%%%%%%%%%%%%%%%%%%%%%%%%%%%%%%%%%%%%%%%%%%%%%%%%%%%%
\section{System Description}\label{s:expSetup}

\subsection{Experimental Setup}
\begin{figure}[b]
  \begin{center}
   \includegraphics[scale=1]{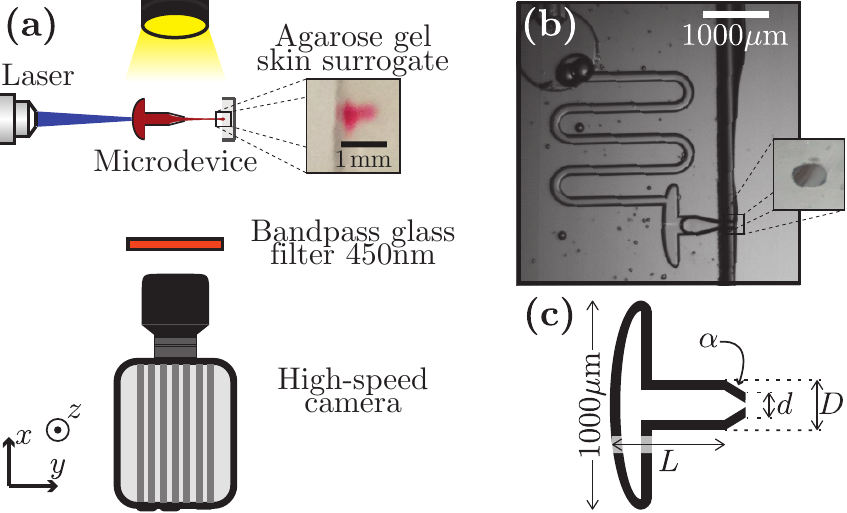}
  \end{center}
  \vspace{-0.3cm}
  \caption{(a) Schematic of the experimental setup: A laser is focused at the bottom of a microfluidic device using a microscope objective. As a result, the bubble and jet are formed and are recorded using an ultra high-speed camera. The inset shows the agarose gel holder and outcome of one jet injection. (b) Top view image of the microdevice: The liquid is introduced through the microtubing toward the microchannel, passing through a spiral tube to prevent the liquid from moving backward. The inset shows the microdevice nozzle. The bonding plane is not visible due to the almost perfect sealing provided by anodic bonding of glass wafers. (c) Schematic of the microdevice: The channel diameter is kept constant at $D=240~\mu$m and, for tapered nozzles ($\alpha>0$), the nozzle diameter is $d=120~\mu$m.}
  \label{fig:ExpSetup}
\end{figure}

The experimental setup consists of a transparent glass -Borofloat$^{\tiny\textregistered}$- microdevice which is partially filled with a water solution containing a red dye, see Fig.~\ref{fig:ExpSetup}. A laser diode (Roithner LaserTechnik), with a wavelength $\lambda=450$~nm and nominal power of 3.5 W, is focused at the microchannel wall opposite to its exit with a 10$\times$ microscope objective. The laser spot size has an ellipsoidal shape, with beam diameters $r_x=33~\mu$m and $r_y=6~\mu$m and variable power P = 400-600~mW. A vapour bubble is formed by the absorption of the laser energy pushing the surrounding liquid through the nozzle and creating a jet that penetrates in agarose gels located in front of the open end of the microchannel at a stand-off distance of 3 mm. The bubble growth and the liquid jet formation are recorded at $8\times10^5$~frames per second (fps) using a high-speed camera (Phantom v2640). The laser light  is blocked from the camera sensor using a notch filter at $\lambda=450$~nm.

\vspace{0.5cm}
\noindent\textbf{\textit{Micro-device design}:} Microfluidic chips were designed and fabricated in glass substrates under cleanroom conditions~\cite{Berrospe}. The microdevice has a microfluidic chamber where the bubbles are created, and is connected to a channel that can be either straight or tapered, all having 100~$\mu$m depth ($z$ direction), see Fig.~\ref{fig:ExpSetup} (a) and (b). The liquid is introduced through the chamber using capillary tubings connected to a precision glass syringe, and controlled by a syringe pump (Harvard PHD 22/2000). We have used three geometrical designs of the tapered channel, with corresponding angles $\alpha=0^\circ,\,14^\circ,37^\circ$, where $0^\circ$ is a straight channel. For tapered nozzles ($\alpha>0$) the nozzle diameter is $d=120~\mu$m and the channel diameter is kept constant at $D=240~\mu$m. The channel length $L$ varies in relation to the taper angle, as shown in Fig.~\ref{fig:ExpSetup}~(c). 

%\begin{figure}[t]
%  \begin{center}
%   \includegraphics[scale=1]{Devices.pdf}
%  \end{center}
%  \vspace{-0.3cm}
%  \caption{(a) Top view image of the microdevice: The liquid is introduced through the microtubing toward the microchannel, passing through a spiral tube to prevent the liquid from moving backward. (b) Picture of the microdevice nozzle. The bonding plane is not visible due to the almost perfect sealing provided by anodic bonding of glass wafers. (c) The three microdevices using in this study are presented. The channel diameter is kept constant at $D=240~\mu$m and, for tapered nozzles ($\alpha>0$), the nozzle diameter is $d=120~\mu$m.}
%  \label{fig:Devices}
%\end{figure}

\vspace{0.5cm}
\noindent\textbf{\textit{Liquid solution}:} In order to maximise the energy absorbed by the liquid from the focused laser, a red dye (Direct Red 81, CAS No. 2610-11-9) was diluted in distilled water at 0.5~\%wt. The corresponding solution has a density$\rho=1000$~kg/m$^3$ (22$^\circ$C), viscosity $\eta=0.91$~mPa$\cdot$s (22$^\circ$C) and surface tension $\sigma=47$~mN/m (22$^\circ$C). They were measured using a density meter (Anton Paar DMA$^\text{TM}$ 35 Ex), a rheometer with cone-plate geometry (Anton Paar MCR 502) and a dynamic contact angle measuring device and tensiometer (Data Physics DCAT 25), respectively. 

\vspace{0.5cm}
\noindent\textbf{\textit{CW laser diode}:} CW lasers can be switched within microseconds with relative ease by switching the driving current. Broad stripe diode lasers can generate  one Watt level output. Since the power is limited by nonlinear (peak power) damage to the facet, the CW power limit is similar to the peak power limit. For these experiments, the laser was switched on for several milliseconds at a power level of 500~mW. The recordings presented in this manuscript are shown starting at 600~$\mu$s. At that point in time, 300~$\mu$J has been delivered. The absorption in the solution ($10^4$/ based on Beer-Lambert's law) is so high that most of the laser energy is absorbed in the first 200~$\mu$m in the solution. The laser is focused with a transverse spot size of 33 $\times$ 6~$\mu$m$^2$. The 300~$\mu$J energy delivered into $4\cdot10^4~\mu$m$^3$  with an absorption length of 200~$\mu$m is enough to heat the (largely water) fluid of that volume well over a 100$^\circ$C. Therefore, the boiling point is quickly reached and a bubble is formed, starting at the chamber interface. If we assume that absorption continues, from room temperature at 20$^\circ$C to reach the boiling point at 100$^\circ$C, a volume of $4.60\cdot10^5\mu$m$^3$ can be heated, which corresponds to a sphere with a diameter of nearly 100~$\mu$m.

%%%%%%%%%%%%%%%%%%%%%%%%%%%%%%%%%%%%%%%%%%%%%%%%%%%%%%%%%%%%%%%%%%%%

%%%%%%%%%%%%%%%%%%%%%%%%%%%%%%%%%%%%%%%%%%%%%%%%%%%%%%%%%%%%%%%%%%%%
%%%%%%%%%%%%%%%%%%%%%%%%%%%%%%%%%%%%%%%%%%%%%%%%%%%%%%%%%%%%%%%%%%%%
\section{Results}\label{s:Analysis}
\subsection{Growth and collapse of jet-producing bubbles}
\begin{figure}[h!]
  \begin{center}
   \includegraphics[scale=1]{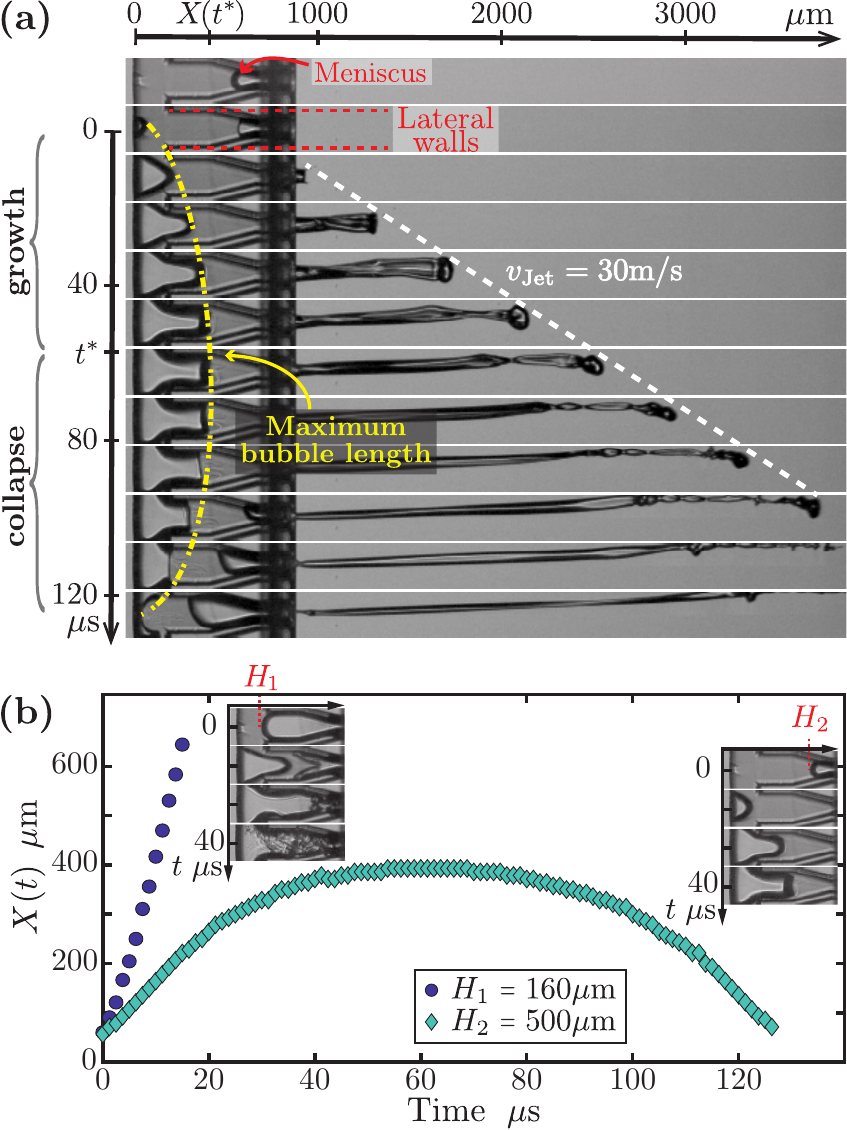}
  \end{center}
  \vspace{-0.3cm}
  \caption{The bubble length over $X(t)$ time is represented for the device taper angle $\alpha=14^\circ$. (a) An image sequence shows the growth and collapse of the bubble inside the microfluidic device, for $H=380~\mu$m. The white dashed line corresponds to the position of the jet tip and its slope represents the jet speed $v_\text{Jet}=30$m/s. As the bubble grows the liquid is guided out to the channel, the growth and collapse of the bubble is represented by the dot-dashed yellow line. The bubble reaches its maximum length $X(t^*)$ at the time $t^*\sim55\mu$s. (Multimedia view) (b) The same conditions with initial meniscus position $H_1=160\mu$m and $H_2=500\mu$m are plotted against time: for $H_{1}$, all the liquid is vaporised or expelled from the device, while for $H_{2}$, the bubble collapses.} 
  \label{fig:BubbleGrowth}
\end{figure}

Fig.~\ref{fig:BubbleGrowth}(a) shows a typical cavitation-induced jet formation for a microdevice with taper angle $\alpha=14^\circ$. As the bubble grows the fluid is directed out to the channel and a liquid jet is formed, see Fig.~\ref{fig:BubbleGrowth}(a). The displacement of the jet tip increases linearly over time, which allows the calculation of the jet speed from the slope of the position curve (dashed line). As can be seen, the laser-created bubble grows from the entry point of the laser beam, until the bubble gas-liquid interface reaches the lateral walls. From then on, the bubble grows further in the axial direction with an elongated pancake shape along the walls of the cell and constrained by the wall onto which the laser is focused~\cite{ohl_acoustic_2016, xiong_droplet_2015, shen_dynamics_2014, zhu_pancake_2016,Brujan} and a constant cross-sectional area, as shown in Fig.~\ref{fig:BubbleGrowth}~(a). The growth and collapse of the bubble depend on the initial amount of liquid contained in the channel, i.e. the initial retracted meniscus position $H$ with respect to the bottom of the microcell \cite{Peters, Tagawa}. 

Depending on the total liquid volume and laser energy input, two scenarios have been observed: the first occurs when all the liquid is vaporised (H$_{1}$), while the second exhibits a collapse phase with complex after-bounces and secondary cavitation close to the bottom of the cell, see Fig.~\ref{fig:BubbleGrowth}~(b). In the second scenario (H$_{2}$), the bubble grows and collapses quasi-symmetrically in time, in contrast to the work by Zwaan \textit{et al.} \cite{Zwaan} in non-confined conditions, and Sun \textit{et al.} \cite{sun_can_dijkink_lohse_prosperetti_2009} in microtubes in which the bubble growth is always faster than bubble collapse. We attribute this observation to viscous dissipation effects given by boundary layer development during expansion and collapse, which are more pronounced in our lower energy experimental conditions.

%\begin{figure}[h!]
%  \begin{center}
%   \includegraphics[scale=1]{ImageSequenceChip08.pdf}
%  \end{center}
%  \vspace{-0.3cm}
%  \caption{A sequence of images showing the liquid jet formation, corresponding to a device with angle $\alpha=14^\circ$. The white dashed line corresponds to the position of the jet tip and its slope represents the jet speed $v_\text{Jet}=30$m/s. As the bubble grows the liquid is guided out to the channel, the growth and collapse of the bubble is represented by the dot-dashed yellow line. The bubble reaches its maximum length $X(t^*)$ at the time $t^*\sim55\mu$s. (Multimedia view)}
%  \label{fig:JetSeq}
%\end{figure}

%%%%%%%%%%%%%%%%%%%%%%%%%%%%%%%%%%%%%%%%%%%%%%%%%%%%%%%%%%%%%%%%%%%%
\subsection{Dimensionless numbers in the jet dynamic}
The microdevice geometries used in this study have the same depth $\ell=100~\mu$m, with specific taper angles ($\alpha=0^\circ,\,14^\circ,\,37^\circ$), and nozzle diameters ($d=120,\,240\mu$m). Considering $\ell$ as the characteristic length of our system, we can calculate 
the dimensionless Ohnesorge number of the jet as
\begin{equation}
\text{Oh}=\frac{\eta}{\sqrt{\rho\sigma\ell}}=0.015,
\end{equation}
\noindent where $\rho$ and $\sigma$ are the density and surface tension of the liquid. This value is much smaller than 1, showing that the jet dynamics is dominated by inertial and surface tension forces. In terms of the jet breakup behaviour, this value is smaller than the critical Ohnesorge number proposed by Reis and Derby, Oh$^*$\hspace{-0.05cm}=0.1, for stable drop formation \cite{reis_derby_2000}.
The minimum dimensionless Weber number of the liquid, related to the minimum jet speed reached in the experiments $v^\text{min}_\text{Jet}\simeq20~$m/s, is
\begin{equation}
\text{We}^\text{min}=\frac{\rho (v^\text{min}_\text{Jet})^2\ell}{\sigma}=850,
\end{equation}
\noindent and the corresponding Weber number in the surrounding gas (air) is
\begin{equation}
\text{We}^\text{min}_\text{gas}=\frac{\rho_\text{air}}{\rho}\text{We}^\text{min}=1,
\end{equation}
\noindent where $\rho_\text{air}$ is the air density. Hence, in our experiments, the minimum Weber number observed in the gas is larger than the critical We$_\text{gas}^*=0.2$, to have droplet formation in the Rayleigh breakup regime \cite{LinReitz,lefebvre1988atomization}. 

Jets such as those observed in Figures 3 and 4 are asymmetric and unstable. The asymmetry is caused by the non-axisymmetric shape of the nozzle (inset of Fig. 2(b)). The initial cross-sectional shape of the jet is flattened instead of circular when it leaves the nozzle, and due to the combination of surface tension and inertia there are cross-sectional shape oscillations along the length of the jet. Interestingly, our experimental data falls in the wind-induced regime, but we do not attribute the jet breakup to aerodynamic effects. We hypothesize that the geometrical constraints of the liquid, combined with the use of a CW laser, produce bubble dynamics that induce inertial effects in the breakup events we observed. The cases described above, are represented in Fig.~\ref{fig:OurWork}~(a) where the orange rectangle corresponds to this work.
%\sout{

%%%%%%%%%%%%%%%%%%%%%%%%%%%%%%%%%%%%%%%%%%%%%%%%%%%%%%%%%%%%%%%%%%%%
\subsection{Jet penetration characteristics}

\begin{figure}%[htb!]
  \begin{center}
   \includegraphics[scale=1]{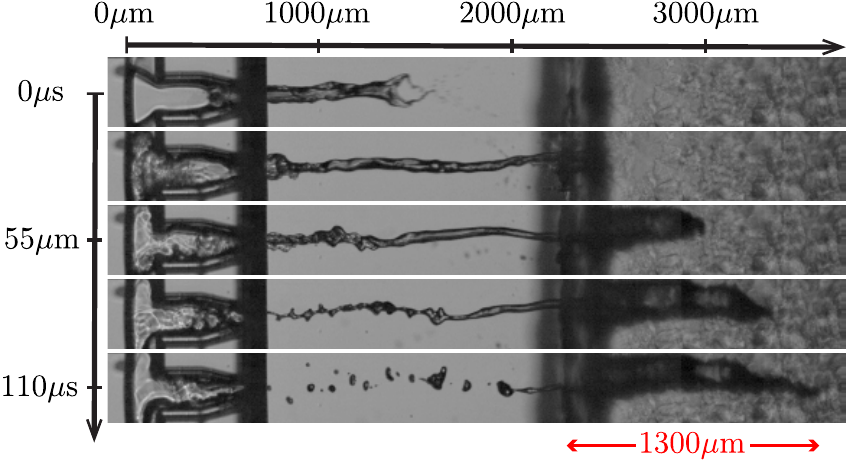}
  \end{center}
  \vspace{-0.3cm}
  \caption{Images sequence showing the liquid jet penetrating in an agarose hydrogel $1\%$wt, corresponding to the device with angle $\alpha=14^\circ$. For a jet speed $v_\text{Jet}=48$m/s, the corresponding depth penetration is $L_m\simeq1300\mu$m.  Note that there is some splash-back of liquid, and the cavity that was formed in the agarose is wider than the jet diameter. As a result, the volume of delivery $V_D$  is slightly less than 100~\%. (Multimedia view)}
  \label{fig:Injection}
\end{figure}

As presented in Fig.~\ref{fig:OurWork}, our experimental conditions correspond to jet powers between 50~mW and 6~W, in $1~\%$wt (OmniPur agarose, CAS No. 9012-36-6.), and the delivered volume percentage amounts to $V_D \sim 100~\%$ with  penetration depth ranging up to $L_m\simeq2$~mm, as shown in Fig.~\ref{fig:Injection}.  We observed that cavities in gel substrates are often wider than the jet, which has been attributed to air entrainment during injection~\cite{rohilla2019vitro}.

\begin{figure*}[t]
  \begin{center}
   \includegraphics[scale=1]{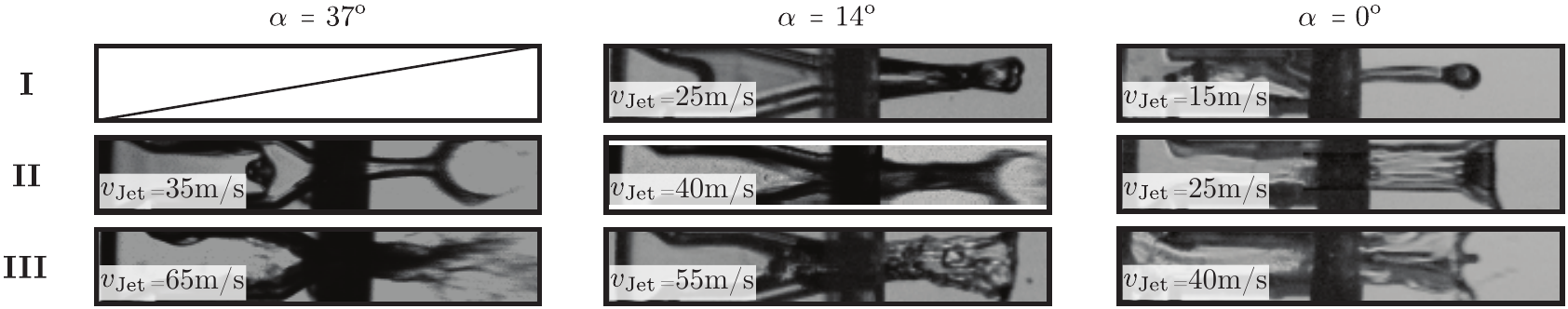}
  \end{center}
  \vspace{-0.3cm}
  \caption{The distinctive jet regimes observed in the experiments for three taper angles. In regime I, the jet tip has a semi-spherical tip shape, in II, which can be considered a transition regime, the jet tip attains a fork shape and in III the liquid jet is a high Reynolds number turbulent jet, Re$>$5000, which could lead, for example, to atomisation or spray.}
  \label{fig:Regimes}
\centering
\end{figure*}
%\FloatBarrier

%%%%%%%%%%%%%%%%%%%%%%%%%%%%%%%%%%%%%%%%%%%%%%%%%%%%%%%%%%%%%%%%%%%%

\subsection{Experimental and numerical results comparison of jet regimes}\label{ss:ExpNum}

Depending on the jet speed, three different jet regimes are observed related to its shape, as shown in Fig.~\ref{fig:Regimes}. In regime I, a focused jet is formed, the jet tip has a semi-spherical tip shape which is highly desired for ink-jet printing~\cite{Hue,Park2007,Hongming} and its diameter is equal to or smaller than the nozzle diameter $d$. For the taper angle $\alpha=37^\circ$, the first regime was not experimentally observed, suggesting that a such pronounced taper angle does not form a focused jet.  Recently, it was shown that, for larger volume jet injectors, the orifice radius does not necessarily control imparted kinetic energy and the jet does not always remain collimated~\cite{rohilla2019vitro}.

\begin{figure}%[h!]
  \begin{center}
   \includegraphics[scale=1]{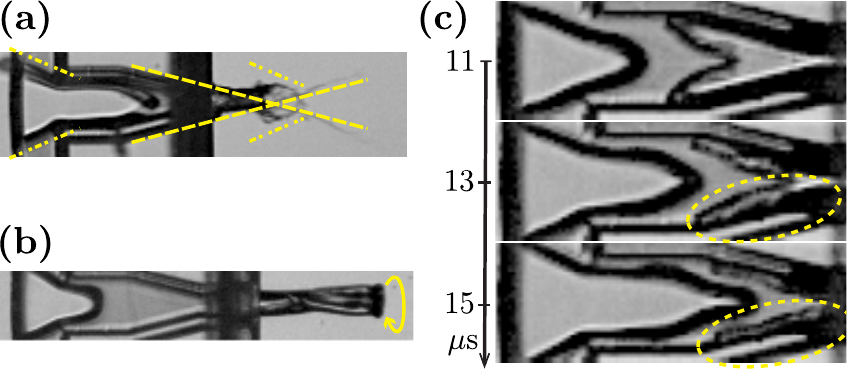}
  \end{center}
  \vspace{-0.3cm}
  \caption{(a) The fork-shaped jet tip is represented where the dashed lines show the extrapolation of the taper walls, and the dotted-dashed lines indicate the extrapolation of the deformed bubble angle. This image is a zoom-in at t=15$~\mu$s of the image sequence in Figure~\ref{fig:Injection}. (b) An example of the swirling jet, corresponding to the image sequence at t=16$~\mu$s in Fig.~\ref{fig:Injection}. (c) The image sequence shows the trapped air process as the meniscus moves forward (Multimedia view).}
  \label{fig:CrossFlow}
\end{figure}

For all device geometries, we observed a jet speed threshold value (transition between regime I and II) after which the jet tip attains a fork shape. This fork shape characteristics of regime II, we attribute it to the tapering and the growing bubble deformation as it enters the channel. Initially, as the jet is moving out of the nozzle, two separated flow threads are created due to the liquid-wall viscous effect in the taper. The flows have a prescribed angle $\alpha$ corresponding to the taper angle and cross each other outside of the device, as shown by the yellow dashed line in Fig.~\ref{fig:CrossFlow}~(a). Then, when the bubble reaches the main channel, another crossing flow is created, this time with a given angle corresponding to the deformed bubble, as shown by the dot-dashed line in Fig.~\ref{fig:CrossFlow}~(a). Moreover, the crossing flows due to the taper, not only induce the fork shape, but also the formation of a swirling jet, as shown in Fig.~\ref{fig:CrossFlow}~(b), which is observable for all regimes, taper angles and jet speeds. Though it is not possible with our current tools to quantify the effect of small defects at the bonding plane between the glass wafers that form the microchannel, these may indeed lead to disturbances in the liquid as it flows through the nozzle exit.

Finally, for sufficiently high jet velocities (regime III) as the liquid flows through the nozzle, wall-bounded air is entrapped in the liquid, and lead to a high Reynolds number turbulent jet, Re$>$5000, as shown by the dashed-ellipses in Fig.~\ref{fig:CrossFlow}~(c), and in some cases atomisation or spray~\cite{TurbulentJets, lefebvre1988atomization}. 
 
The fork shaped jet tip is not desired for specific applications, such as needle-free injections where a high pressure to penetrate the skin is necessary and thus, a circular cross-section with the smallest possible diameter gives the best results. However, due to its high speed, the jet penetrates agarose slabs up to a depth of $\sim$2 mm, see Fig.~\ref{fig:Injection}. 

\begin{figure}[b]
  \begin{center}
   \includegraphics[scale=1]{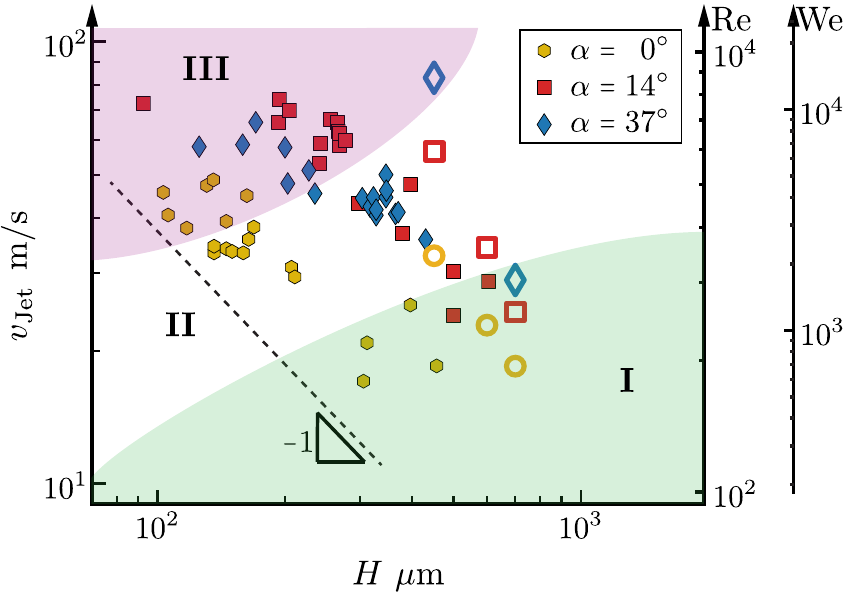}
  \end{center}
  \vspace{-0.3cm}
  \caption{The jet speed versus the initial meniscus position $H$, with different symbols representing the taper angle of the microdevice. The solid symbols represent experimental data and the open symbols numerical simulation results (each symbol represents a single experiment). The dashed line corresponds to a power law with exponent -1, as suggested by Ory \textit{et al.}~\cite{ory2000growth} and Peters \textit{et al.}~\cite{Peters}, and is plotted to compare with inviscid flow theory. The coloured areas  represent to the three regimes described in Figure~\ref{fig:Regimes}.}
  \label{fig:SpeedVsMeniscus}
\centering
\end{figure}

The experimentally measured  jet speed for the three microdevices taper angles is plotted with respect to the initials meniscus position $H$ in Fig.~\ref{fig:SpeedVsMeniscus} solid symbols. The figure also includes the mentioned regimes (I, II and III); regime I is the green area corresponding to higher H and lower jet speed, regime III has the lowest H and highest jet speed, and regime II is the intermediate regime. A large scatter in the data is observed, which we attribute to the initial condition in the nozzle and the walls, e.g. the presence of water droplets from a previous jet ejection \cite{MORADIAFRAPOLI2017110}. The jet speed is observed to increase under two specific conditions. One is determined by the geometry, more specifically the velocity increases once the nozzle is tapered. Our results show that for a given filling factor, $v_\text{Jet}$ can increase up to 200\% compared to the straight channel device. The tapered channel helps to focus the liquid causing an increase in the jet speed.  However, it is striking that the experimental jet speeds for the two tapered channels are comparable. This observation suggests that there is an optimum taper angle after which the maximum jet speed does not increase anymore, and might even decrease. In an attempt to explain this result, we used numerical simulations to obtain more detailed information on the jet formation under conditions comparable to those in the experiments (see Section E). 

The second condition that affects the jet speed is related to the initial meniscus position that is represented by the filling factor. For lower $H$ values higher velocities are reached, because the growing bubble has to displace less mass with the same amount of input energy from the laser. The relationship between $v_\text{Jet}$ and $H$ can also be determined by quantifying the pressure changes in the system, based on the assumptions in literature \cite{ory2000growth,Peters}. We assume a bubble growing in a quasi one-dimensional direction, and write a simplified Navier-Stokes equation for an inviscid and incompressible flow as:
\begin{equation}
\frac{\partial v}{\partial t}=-\frac{1}{\rho}\frac{\partial p}{\partial x},
\label{Eq:Navier}
\end{equation}

\noindent where $v$ is the local velocity of the fluid, $\rho$ the density of the liquid and the pressure gradient is the ratio between the given initial pressure and the meniscus position $\partial p/\partial x=\Delta p/H$. Thus, integrating equation~\ref{Eq:Navier} over time $\Delta t$, we can write the velocity of the free surface, i.e. $v_\text{Jet}$, after the bubble nucleates as: 
\begin{equation}
v_\text{Jet}=\frac{\Delta p \Delta t}{\rho H}.
\label{Eq:VjetvsH}
\end{equation}
The power law $v_\text{Jet}\sim H^{-1}$ is represented by the dashed line in Fig.~\ref{fig:SpeedVsMeniscus}, as well as the three regimes described at the beginning of this section.

\begin{figure}[t]
 \begin{center}
  \includegraphics[scale=1]{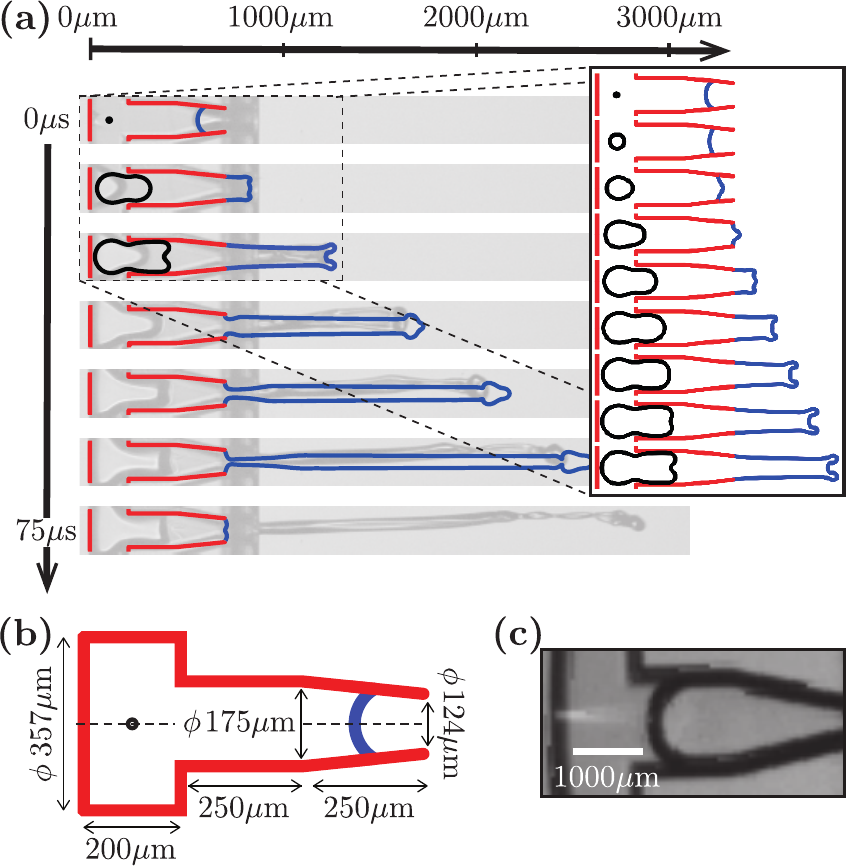}
 \end{center}
 \vspace{-0.3cm}
 \caption {Jet formation and pinch-off for a BI simulation in blue and red, with $\Delta p$~=~10~bar and $\Delta t=10\mu$s. In overlay, we compare the simulation with the experimental results shown in Fig.~\ref{fig:BubbleGrowth}~(a). After $t=33\mu$s the BI code removed the bubble because of instabilities on its surface caused by the proximity of the wall. The inset shows in more detail the bubble growth within the first 25~$\mu$s. (b) Numerical setup in the axisymmetric BI code, consisting of a channel wall (red), a meniscus (blue), and a bubble surface with an initial diameter of 175 $\mu$m (black). (c) A picture of its analogous experimental device ($d=120~\mu$m$,\,\alpha=14^\circ$), where the laser spot is seen as a lighter area. (Multimedia view) }
 \label{fig:JetSeqSims}
\end{figure}

%%%%%%%%%
\subsection{Numerical simulations of jet formation}

The jet formation was simulated using an axisymmetric boundary integral (BI) code, in which the liquid flow is assumed to be irrotational, incompressible, and inviscid \cite{oguz1993dynamics, power1995boundary, bergmann2009controlled, Geckle2009high-speed}. The numerical configuration was adapted from previous studies on jet formation by pulsed laser-induced cavitation~\cite{Peters}, and consists of a nozzle wall, a meniscus, and a bubble surface, see Fig.~\ref{fig:JetSeqSims}~(a). A picture of the analogous experimental device is shown in Fig.~\ref{fig:JetSeqSims}~(b). We use BI to provide a qualitative comparison of the jet shapes, and we limited our analysis to feature extraction.  

%\begin{figure}[b]
%\includegraphics[scale=1]{NumericalSetupV2.pdf}
%\caption{(a) Numerical setup in the axisymmetric BI code, consisting of a channel wall (red), a meniscus (blue), and a bubble surface with an initial diameter of 175 $\mu$m (black). (b) A picture of its analogous experimental device ($d=120~\mu$m$,\,\alpha=14^\circ$), where the laser spot is seen as a lighter area.}\label{fig:NumSetup}
%\end{figure}

Since the BI code is axisymmetric, and the experimental setup was not, the radial dimensions of the numerical configuration were calculated based on the cross-sectional areas of the experimental setup. First, the radius of the chamber and the straight section of the nozzle were chosen such that the cross-sectional area of each section equalled that of the corresponding section in the experimental setup. Secondly, the radius was calculated at the start and end of the tapered section of the nozzle based on the cross-sectional area at these positions in the experimental setup. In between the start and end position the radius was calculated based on linear interpolation between these two radii.

The contact angle $\theta$ was measured with respect to the axis of symmetry, i.e.\ independent of the inclination of the nozzle wall, to avoid instabilities on the meniscus shape during transition of the contact line from the straight to the tapered section of the nozzle, and {\it vice versa}. The contact line dynamics of the meniscus were modelled using the contact angle hysteresis concept, with a receding contact angle $\theta_{r}$ and an advancing contact angle $\theta_{a}$. The contact line was pinned in the case where $\theta_{r} < \theta < \theta_{a}$, it was moving to have $\theta = \theta_{r}$ for $\theta < \theta_{r}$, and it was moved to have $\theta = \theta_{a}$ for $\theta > \theta_{a}$. $\theta_{r}$ and $\theta_{a}$ were set to the maximum angle away from 90$^{\circ}$ at which the meniscus motion near the wall remained stable during the simulations, i.e. $\theta_{r}$~=~72$^{\circ}$ and $\theta_{a}$~=~108$^{\circ}$. 

The bubble's initial radius was set to 10\% of the nozzle radius, placed in the centre of the chamber, and not at the wall as occurs in the experiments. This was meant to prevent a premature ending of the simulation due to instabilities of the bubble surface. A rectangular wave pressure pulse was applied to the bubble at the start of the simulation to mimic the pressure evolution in the bubble during the experiment, which is driven by the sudden phase-change after liquid heating and by the rapid cooling of the vapour during bubble expansion \cite{yuan1999growth, yuan1999pumping, ory2000growth, sun_can_dijkink_lohse_prosperetti_2009, Peters}. The bubble growth and jet formation in the simulation were matched to that of the experiment by tuning the amplitude $\Delta p$ and duration $\Delta t$ of the rectangular pressure pulse.

%%%%

An example of a simulation qualitatively reproducing the observed jet formation for a device with angle $\alpha=14^\circ$ and jet speed $v_\text{Jet}=34$~m/s (corresponding to Fig.~\ref{fig:BubbleGrowth}~(a)), is shown in Fig.~\ref{fig:JetSeqSims}. The differences in the timescales of simulations and experimental observations are evident. The parameters $\Delta p$ and $\Delta t$ of the rectangular wave pressure pulse were adjusted to match the bubble and jet characteristics in the simulation. The jet formation was tuned mainly through changes in the $\Delta p$ value, while the maximum bubble size was tuned through $\Delta t$. In this case, the jet formation is similar to that observed in the experiment, however, the bubble growth speed only qualitatively match experimental data. 

\begin{figure}[h!]
  \begin{center}
   \includegraphics[scale=0.9]{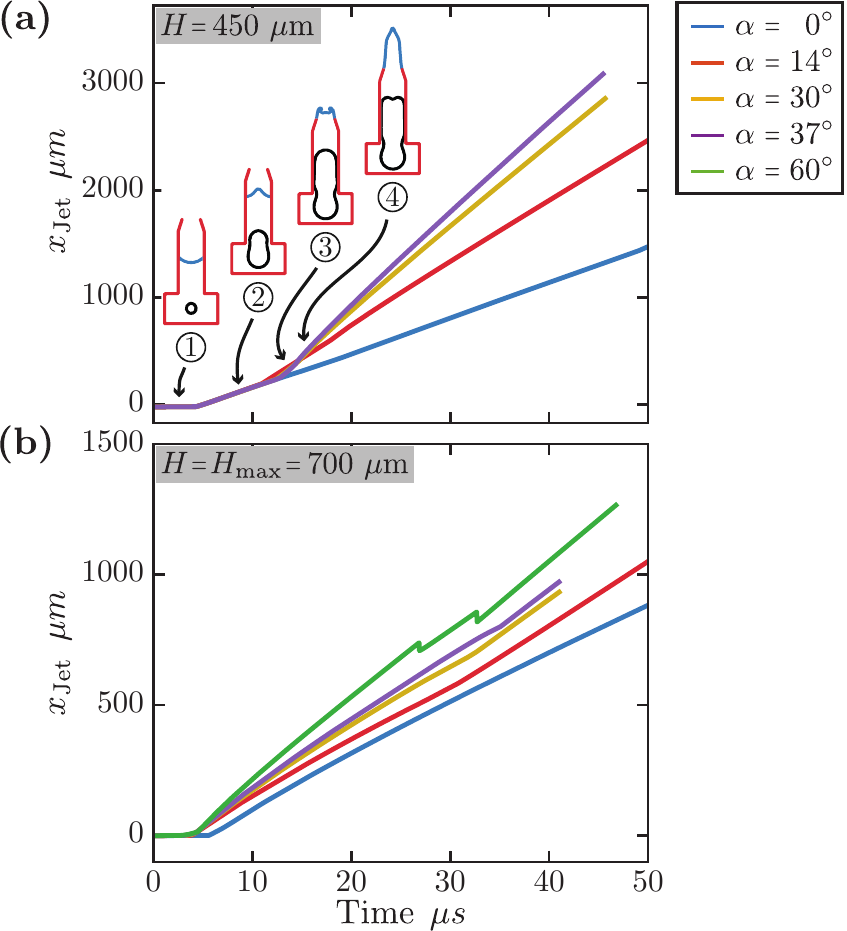}
  \end{center}
  \vspace{-0.3cm}
  \caption{Jet tip position versus time for tapered angles between $\alpha=0^\circ$ and $\alpha=60^\circ$ is plotted in two cases: (a) an initial meniscus position H=450~$\mu$m and (b) a fully filled channel, H=700~$\mu$m. In case (a), three changes in the slope of the position curve are observed due to the mechanisms presented in the insets: \textcircled{\footnotesize{1}} contact line is still fixed; \textcircled{\footnotesize{2}} central jet takes over the lead; \textcircled{\footnotesize{3}} toroidal (fork shaped) jet takes over the lead; \textcircled{\footnotesize{4}} toroidal jet collapsed into a central jet. The jet tip position corresponding to $\alpha=60^\circ$ is not available due to BI code limitations, refer to the main text for more details.}
  \label{fig:DataSims}
\end{figure}

Fig.~\ref{fig:DataSims} shows the jet tip position versus time for tapered angles between $\alpha=0^\circ$ and $\alpha=60^\circ$, from numerical simulations, in two different situations: a partially ($H=450~\mu$m) and a fully filled ($H_\text{max}=700~\mu$m) device. For the partially filled case, the curve changes in slope three times. As shown in the insets of Fig.~\ref{fig:DataSims}~(a), initially the contact line is the maximum position, then a central jet takes over the lead at the first change in slope, subsequently a toroidal (fork shaped) jet takes over the lead during the second change in slope, and finally the toroidal jet collapses into a central jet. The first central jet is due to flow focusing at the concave meniscus~\cite{Peters}. The toroidal jet is the result of a relative high velocity of the meniscus close to the wall, which is amplified by the nozzle taper. The fact that this effect is largely suppressed when the microdevice is fully filled, as shown in Fig.~\ref{fig:DataSims}(b), suggests that the high velocity of the meniscus at the wall is a result of local inhomogeneous flow due to the expanding  and deforming bubble. The smaller changes in slope in Fig.~\ref{fig:DataSims}(b) are due to shape deformations of jet head droplet, while the discontinuities for $\alpha=60^\circ$ are due to the pinch-off of small satellite droplets from the main head droplet. This detailed jet position in time cannot be observed in the experimental data, because of the limited temporal and spatial resolution.

\begin{figure}[!h]
  \begin{center}
   \includegraphics[scale=1]{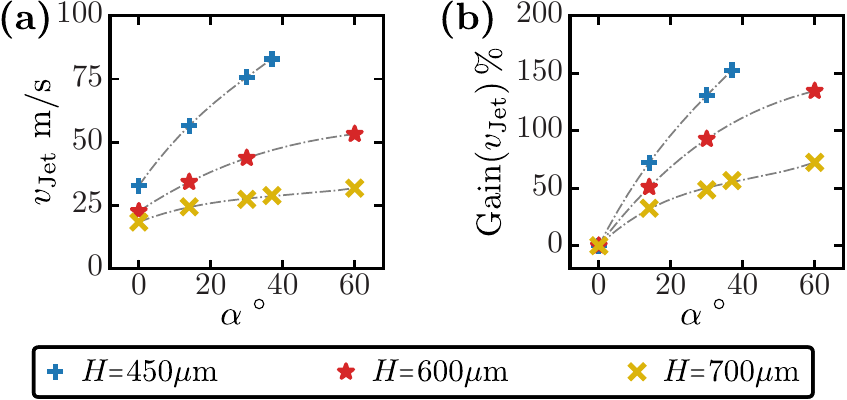}
  \end{center}
  \vspace{-0.3cm}
  \caption{(a) The jet speed $v_\text{Jet}$ and (b) the percent speed gain $\text{Gain}(v_\text{Jet})$, from numerical simulation data are plotted versus the tapered angle. The symbols correspond to the different initial meniscus position $H$, from a half filled device $H=450\mu$m to a fully filled device $H=700\mu$m. The dashed-dotted lines serves as a visual guide.}
  \label{fig:DataSimsSummary}
\end{figure}

In order to compare the experimental data with simulations, we extracted the jet speed $v_\text{Jet}$ for each simulation by calculating the average jet speed at the steady state part of jet tip position curve. The obtained jet speeds are plotted in Fig.~\ref{fig:SpeedVsMeniscus} open symbols and compared with the experimental results. The jet speed in the simulations is higher than that in the experiments. Furthermore, in the simulation data no optimum taper angle is visible; the jet speed increases with increasing taper angle. This is even more clearly visible Fig.~\ref{fig:DataSimsSummary}, which shows for all simulations the jet speed, $v_\text{Jet}$, and the percent jet speed gain, Gain($v_\text{Jet}$), as function of the taper angle~$\alpha$ and initial meniscus position~$H$. $v_\text{g}$~is the ratio of the jet speed to the jet speed for that initial meniscus position in a nozzle without taper. The jet speed increases with the taper angle without reaching an optimum. This is in disagreement with that observed experimentally in Fig.~\ref{fig:SpeedVsMeniscus}~(b), where there is an optimum taper angle required to obtain higher jet speeds between $\alpha\sim14^\circ$ and $\alpha\sim37^\circ$. This difference is likely due to the absence of viscous dissipation in the BI simulations, due to the axisymmetric setup, and due to the simplified bubble pressure model.

Finally, it will be interesting to have experiments and modelling, using finite-element models such as ANSYS-Fluent, where the microfluidic chip is manufactured from materials less stiff than glass, and to account for short-lived pressure impulse effects due to material compliance. The compliance can help tune to the eigenfrequency of the system, which then can lead to a better control of the breakup and length of the jet. Such strategies are typically employed in inkjet printing. Experiments using more compliant materials, including plexiglass and PDMS, will allow a more detailed insight in the role of wall compliance. 

%%%%%%%%%%%%%%%%%%%%%%%%%%%%%%%%%%%%%%%%%%%%%%%%%%%%%%%%%%%%%%%%%%%%
%%%%%%%%%%%%%%%%%%%%%%%%%%%%%%%%%%%%%%%%%%%%%%%%%%%%%%%%%%%%%%%%%%%%
\FloatBarrier
\section{Conclusions}\label{s:Conclusions}

We have elucidated jetting phenomena induced by thermocavitation in microfluidic devices keeping constant CW-laser parameters such as laser wavelength, beam spot size shape and power. We also kept constant the liquid properties viscosity and density -- Newtonian liquids--, and selected specific geometrical designs of the microfluidic device that determine surface properties of the material containing the liquid such as surface wettability and roughness. The influence on the velocity and shape of the jets produced by changes in the taper angle of the nozzle in the experiments were compared with a numerical model. 

Three regimes were observed related to the jet tip shape: I) semi-spherical tip, II) fork shaped tip, and III) turbulent jet. The jet speed was observed to increase with a decreased liquid volume in the microdevice and with an increased taper angle. Moreover, we laid out that for tapered geometries, we observed the higher jet speeds.

As a follow up we plan a comparison of experiments with more complex numerical models taking into account the viscous dissipation, and the asymmetrical cross-sectional area in the microdevice. The first choices are using Gerris and Basilisk~\cite{popinet2003gerris,popinet2009accurate,popinet2018numerical}. Future experiments will cover other power settings, and changing the fluidic properties, which will allow us to operate in other parameter spaces beyond the wind induced regime.

\section{Supplementary Information}
See Supplementary material for the experimental and modelling videos related to specific figures in the manuscript.

\section{Acknowledgements}
We would like to thank Stefan Schlautmann and Frans Segerink for their technical support during fabrication and optical setup construction. We also thank the James W. Bales from the MIT Edgerton's centre for the access to the Phantom high-speed camera and illumination.  A.F and M.V acknowledge the program High Tech Systems and Materials (HTSM) with project number 12802. D.F.R. acknowledges the recognition from the Royal Dutch Society of Sciences (KHMW) that granted the Pieter Langerhuizen Lambertuszoon Fonds, 2016.

%%%%%%%%%%%%%%%%%%%%%%%%%%%%%%%%%%%%%%%%%%%%%%%%%%%%%%%%%%%%%%%%%%%%
%%%%%%%%%%%%%%%%%%%%%%%%%%%%%%%%%%%%%%%%%%%%%%%%%%%%%%%%%%%%%%%%%%%%
\FloatBarrier

\bibliographystyle{ieeetr}
%\bibliography{TaperAngleEffect}{}

\end{document}